\documentclass[useAMS,usenatbib,usegraphicx]{mn2e}
\usepackage{epsfig}
\usepackage{amsmath} %need for line break in equ 3
\usepackage{rotating}           % for sideways tables/figures
\usepackage{color}     
\usepackage{graphicx}
\usepackage{times}
\usepackage{upgreek} % for up $\uptau$ 
\def\kms {\hbox{${\rm km\ s}^{-1}$}}
\def \HI {H{\sc \,i}}

\def\lapp{\ifmmode\stackrel{<}{_{\sim}}\else$\stackrel{<}{_{\sim}}$\fi}
\def\gapp{\ifmmode\stackrel{>}{_{\sim}}\else$\stackrel{>}{_{\sim}}$\fi}

\title[Classification of redshifted \HI\ 21-cm absorption]{Intervening or associated? Machine learning classification of redshifted \HI\ 21-cm absorption}

\author[S. J. Curran]{S. J. Curran\thanks{Stephen.Curran@vuw.ac.nz}\\
School of Chemical and Physical Sciences, Victoria University of Wellington, PO Box 600, Wellington 6140, New Zealand}

\begin{document}

 \date{Accepted ---. Received ---; in original form ---}

\pagerange{\pageref{firstpage}--\pageref{lastpage}} \pubyear{2021}

\maketitle

\label{firstpage}

\begin{abstract}
  In a previous paper we presented the results of applying machine learning to classify whether an
  \HI\ 21-cm absorption spectrum arises in a source intervening the sight-line to a more distant
  radio source or within the host of the radio source itself. This is usually determined from an
  optical spectrum giving the source redshift.  However, not only will this be impractical for the
  large number of sources expected to be detected with the Square Kilometre Array, but bright
  optical sources are the most ultra-violet luminous at high redshift and so bias against the
  detection of cool, neutral gas. Adding another 44, mostly newly detected absorbers, to the
  previous sample of 92, we test four different machine learning algorithms, again using the line
  properties (width, depth and number of Gaussian fits) as features. Of these algorithms, three  gave a
  some improvement over the previous sample, with a logistic regression model giving the best
  results.  This suggests that the inclusion of further training data, as new absorbers are detected,
  will further increase the prediction accuracy above the current $\approx80$\%.  We use the
  logistic regression model to classify the $z_{\text{abs}}=0.42$ absorption towards PKS\,1657--298
  and find this to be associated, which is consistent with a previous study which determined 
  $z_{\text{em}}\approx 0.42$ from the $K$-band magnitude--redshift relation. %212
\end{abstract}

\section{Introduction} 
\label{intro}

In absorption, the $\lambda=21$~cm transition of neutral hydrogen (\HI) traces the cool neutral gas in galaxies.
Although the reservoir for all star formation, at redshifts of $z_{\text{abs}}\gapp0.1$ (look-back times of
$\gapp1$~Gyr) there are only 142 \HI\ 21-cm absorption systems currently known. Of these, 56 are due to a quiescent
galaxy {\em intervening} the line-of-sight to a more distant radio continuum source and 82 are {\em associated} with the
host of the radio source itself.  

With its large instantaneous bandwidth, forthcoming surveys with the {\em Square Kilometre Array}
(SKA) are expected to vastly increase the number of redshifted 21-cm absorption systems, which will
give an unbiased census of the distribution and abundance of the star-forming reservoir over cosmic
history. For example, observations of the Lyman-$\alpha$ transition in intervening systems show a
flat redshift evolution of the cosmological mass density of \HI\ (e.g. \citealt{cmp+17}), which is
at odds with the $z\sim2$ peak in the star formation density \citep{hb06,bwc13}. However, 
observations of 21-cm absorption suggest that the fraction of cold neutral gas exhibits a similar evolution as the star
formation density \citep{cur19}, thus providing a solution to the long-standing issue of the mismatch
between the star-forming activity and the fuel supply.

Instantaneous coverage of the whole $z = 0 -1$ redshift space is already possible with the
{\em Australian Square Kilometre Array Pathfinder} (ASKAP, \citealt{asm+15a}), yielding several new
detections \citep{asm+15,amm+16,asd+19,mas+17,sma+20}.
Such spectral scanning is a powerful tool in detecting 21-cm absorption without the need for an
optical spectrum, the requirement of which introduces a bias towards the most luminous and least
dust-obscured sources, rendering the cool neutral gas undetectable \citep{cwm+06,cww+08}. Whereas a
detection of 21-cm absorption will yield a redshift, in the absence of an optical spectrum it is not
known whether this is due to intervening or associated absorption.  For example, it was unknown
whether the $z_{\text{abs}}=0.44$ \HI\ 21-cm absorption detected towards PKS\,B1740--517 with
ASKAP's {\em Boolardy Engineering Test Array} was associated until Director's Discretionary Time on
{\em Gemini-South} confirmed optical emission lines at the same redshift \citep{asm+15}.

Follow-up deep optical spectroscopy will be impractical for a large number of (optically faint)
sources and impossible for those which do not exhibit spectral lines (blazars) or are
optically undetected (e.g. \citealt{ysdh12}). One
possible solution is  the attainment of a photometric redshift via machine
learning, specifically deep learning using neural networks.  Training on optically selected samples
(the {\em Sloan Digital Sky Survey}, SDSS), shows promising results with the validation giving
accurate redshift predictions for quasi-stellar objects (QSOs,
\citealt{ldlr11,bcd+13,dp18,pp18,bsf+20}).  Training on SDSS QSOs can also provide a model which
accurately predicts redshifts for radio selected sources (quasars, \citealt{cmp21}).  The model,
however, requires measurements in nine photometric bands, spanning from the near-infrared to
far-ultraviolet, which may not necessarily be available.

Thus, it would of great value to be able to confidently ascertain the nature of the 21-cm absorption
without the need of an optical spectrum nor extensive multi-band photometry. Previously
\citep{cdda16}, we used machine learning techniques, with the absorption line properties as features (see Table~\ref{t1}), 
to find that the absorber type could be
predicted with $\approx80$\% accuracy. This is very promising for such a small training sample and
so here we add a further 44, mostly new, systems to the 92 previously tested in order to 
further test the potential of this method.

\section{Analysis}

\subsection{The data}

\subsubsection{The sample}

Previously \citep{cdda16}, the sample comprised 98 absorbers of which 53 were associated and 43 intervening.
Checking the literature for detections published since and including absorbers which were missed, gives 136 in total
of which 80 are associated and 56 are intervening (Table~\ref{t1}).\footnote{There is an additional six $0.1 \lapp z_{\text{abs}} \leq0.21$ associated absorbers in \citet{gmmo14}, but since the spectra are unlabelled we cannot use these.} Note that during this process, we found duplicates and errors in the  original data and, accounting for 
these, gives 48 associated and 44 intervening systems. 
As before, we include only absorption redshifts of $z_{\text{abs}}\geq0.1$ in order to: 
\begin{itemize}
\item[--]  Avoid resolved sight-lines, which will introduce a systematic difference between a low and high redshift sample.
\item[--] Minimise dilution of the absorption by 21-cm emission, which becomes an issue  at  $z\lapp0.1$ (e.g. \citealt{rsa+15}).
\item[--] Maintain similar sample sizes between the two classes: If one sample is significantly larger than the other, the 
machine learning will simply favour the larger one based upon its higher probability of occurrence.
\end{itemize}
\begin{table*} 
\centering 
\begin{minipage}{168mm}
  \caption{The features of the $z_{\text{abs}}\geq0.1$ \HI\ 21-cm absorbers not in the \citet{cdda16} sample.  The first
    column gives the IAU name, followed by absorber type (associated/intervening), the absorption redshift,
   the number of Gaussian components fit to the profile, the Full-Width at Zero
      Intensity, the peak observed optical depths, the average offset of the
    components from the weighted mean velocity of the absorption, followed by this
    normalised by the FWZI and the Full-Width Half Maxima of the Gaussians.  The last column
    gives the reference for the spectrum.}
\begin{tabular}{@{}lc l  cc   ccc   r r rrr  l@{}} 
\hline
IAU & Type & $z_{\text{abs}}$   & $n_{\rm g}$  & FWZI    & \multicolumn{3}{c}{$\tau_{\rm peak}$} & $\overline{\Delta v}$ & $\overline{\Delta v}$/FWZI & \multicolumn{3}{c}{FWHM} & Ref.\\
       &           &                 &           &             & ave & max & min                                   &                                  &                            & ave & max & min & \\
        &           &               &                    & [\kms]   &               & &                                               & [\kms]  & & \multicolumn{3}{c}{[\kms]} & \\
\hline
B0003+380  & A & 0.229   & 2 & 59 & 0.032 & 0.053 & 0.011& -8.92 & -0.15 & 19.5 & 30.7& 8.3 &  A18 \\ % 
B0035+22 & A & 0.096 & 2 & 189 & 0.015 & 0.017 & 0.012 & -5.0 & -0.026& 75.7 & 107 & 44.5 & O17\\
J0146--0157 & A& 0.95904 & 3 & 1735& 0.009 & 0.012 & 0.007 & -25.5 &  -0.015& 1019 & 1799 & 234& A19 \\ % flux 1804 mJy
B0202+14 & A & 0.8336 &  3 & 1155 & 0.13 & 0.022 &  0.004 & -71.4 & -0.061 & 194 & 416 & 83.0 & J18\\% 
B0229+0044 & A & 1.2166 &  2& 248 & 0.578 & 0.871 & 0.285 & -19.4 & -0.078 & 60.5 & 75.6 & 45.3 & C20 \\
B0229+0053 & A & 1.1630 &2 & 365 & 0.327 & 0.478 & 0.177 & 34.6 & 0.095 & 84.4 & 94.9 & 73.8 & C20 \\  
B0248+430  & I & 0.3939 & 3 & 52 & 0.160  & 0.191 &  0.117 & 0.57 & 0.011 & 5.35 & 6.31 & 4.75 &  L01 \\
B0834--20  & I & 0.5906   & 3 & 78 & 0.090 & 0.107 & 0.080& -4.73 & -0.060 & 17.3 & 31.4 & 4.08 & S20 \\ 
J0903+1622 & A &  0.1823  &2 &   239 &   0.096 & 0.122 &  0.070 & 29.4  &0.12 &  64.8 & 66.9 & 62.7 &  M17a \\
J0919+0146   & I & 1.27307   & 2 & 223 & 0.006& 0.006&  0.006&  -10.7 & -0.045 & 107& 177& 36.4& D17b\\ 
J0921+6215 & I & 1.10360 & 1 & 21.5 & 0.040& 0.040 & 0.040 & -0.37 & -0.017 & 8.4 & 8.4 & 8.4 & D17b\\ 
B0941--08 & A & 0.2281 & 3 & 380 & 0.0121 & 0.027 & 0.016 & 6.42 &  0.017 & 64.0 & 123 & 16.3 & O17\\
J1013+2448 & A & 0.94959 & 3 & 412& 0.005 & 0.006 & 0.004 & 59.2 & 0.14& 98.6 & 316 & 31.7 & A19\\
J1039+4612 & A & 0.1861 & 1 & 166 & 0.193 & 0.193 & 0.193 & -8.77 & -0.053 & 84.8 & 84.8 &  84.8 &  M17a\\
B1045+35A & A & 0.84644& 4 & 402 & 0.011 & 0.037 & 0.012 &  -8.21&  -0.020& 94.5 & 190 & 17.9 & A19\\
B1147+557  & A & 0.13855 & 3 & 199 & 0.035 & 0.050 & 0.021 & 12.3 & 0.062& 33.2 & 42.1 & 21.5 &   C11\\ 
B1200+045  & A &  1.226   & 2 & 188 & 0.015 & 0.022 & 0.009& 4.4& 0.023 & 36.6 & 49.4 & 23.7 & A18a\\ %1675 mJy 
J1206+6413   & A &  0.3710 & 2 &  189 & 0.006 & 0.011 & 0.002  & 42.0 & 0.22 & 87.6 & 90.5 & 84.6 & V03\\
J1209--2032    & A    & 0.4040  & 2 & 262 & 0.091 & 0.136 & 0.046 & 22.9 & 0.087 & 77.7 & 124 & 31.7 & M20\\
B1221-423    & A   & 0.171442 & 2 & 321& 0.011  &   0.020 & 0.003  & -33.7  & -0.10 & 110 & 163 & 57 & J10\\
B1245--197     & A &  1.2750 & 3 &   983 & 0.007  & 0.015 & 0.002 & 69.2 & 0.070 & 324 & 608 & 91& A18a\\    %8302 mJy
J1255+1817 & I & 0.75761	& 2 & 59 & 0.053 & 0.054 & 0.052 & 1.74 &  0.029 & 17.2& 25.7 & 8.77 & D17b\\
J1327+4326 & I & 0.95421 & 2 & 39& 0.010 & 0.014 & 0.006 & -5.90 & -0.13 & 11.9 & 14.7 & 9.2 & D17b\\
J1342+5110 & I & 1.48815 & 4 & 83 & 0.061 & 0.063 & 0.055 & 2.21 & 0.027 & 12.3 &  24.4 & 1.31 & D17b\\
B1345+12     & A &  0.12174 & 2 & 249 & 0.006 & 0.010 & 0.002 & 27.3 & 0.11 & 95.8 & 121 & 71.0 & G06\\
J1357+0046 & A& 0.797 & 2 & 330 & 0.016 & 0.017 & 0.015& 18.7 & 0.057  & 79.9 & 99.1 & 60.7& Y16\\
J1428+2103 & I & 0.39401 & 2 & 27 & 0.171 & 0.220 & 0.122 & 0.12 & 0.004 & 6.69 & 9.67 & 3.71&  D17a \\
B1456+375  & A & 0.33343& 2 & 31 &  0.218 & 0.227 & 0.210 & -1.50& -0.048 & 7.13 &10.1 & 4.20 & A18b\\ %FLUX 148.4 
J1513+3431  & A & 0.1272  & 1 & 276 & 0.300& 0.300& 0.300 & 7.81 & 0.028 & 146 & 146 & 146 & M17a\\
J1521+5508 & A &  1.0701 & 3 & 90 &  0.010 & 0.012 & 0.008 & 6.09 & 0.067 & 21.2 & 46.5 & 5.39 & D19\\  
J1534+2909 & A & 0.201 & 4 & 464 & 0.029 & 0.038 & 0.011 & -12.4& -0.027 & 79.6 & 122 & 31.3 & M17a\\
B1540--1453 & A & 2.1139    &2 & 253 & 0.063 & 0.103 & 0.024 &    -9.63 & -0.038& 100  &  133 &  67.1 &  G21\\
B1549--79  & A & 0.15010 & 2 & 260& 0.015 & 0.018 & 0.05 &  -8.4& -0.032 & 154 & 271 & 37.5 & M01\\
J1551+0713  & I & 0.32891    &  1 & 12 &  0.097 &  0.097 & 0.097 & -0.48 &  -0.04 &  3.97 & 3.97 & 3.97 & D17a \\
B1555--140   & A & 0.0971 & 2 & 541 &  0.076&  0.079 & 0.072 &  -14.1& -0.026 & 137 & 147 & 127 & C06\\ 
J1602+5243 & A & 0.1057 & 2 & 207 & 0.007 &  0.010 & 0.004 & -27.6 & -0.13 & 53.56 & 58.0 & 49.1 &C11\\ % also in gmmo14
B1610--77  & I & 0.4503  & 3 & 87   &  0.027 & 0.037 & 0.022 & 2.82&  0.032 & 18.5 &  30.0 & 5.9 & S20\\ % 3 better than pub 2
B1645+2230 & A & 0.82266 & 2 & 231 & 0.066 & 0.081 & 0.052 & -3.59 & -0.016& 66.0 & 72.5 & 59.6 & A19\\% 252.8 mJy
B1657--298 & -- & 0.42016 & 2 & 181 & 0.031 & 0.054 & 0.009  &  6.9 & 0.038  & 36.2 & 54.3 & 18.2& M17b\\
J1708+2111 & A & 0.2241 & 2 & 356& 0.110& 0.155 & 0.065 & -38.4 & -0.11& 131 & 185& 78 & M17a\\
B1740--517 &  A & 0.4413   & 3 & 211&  0.072 & 0.173 &   0.008 & 16.9 &   0.080 & 17.3 & 58.0 & 10.7 &  A15 \\
B1829--718   & A & 0.536      & 5 & 481 & 0.037 & 0.025 & 0.001 & 193 & 0.40 & 50.6 & 89 & 22 & G19 \\  % DON'T HAVE
B1954+513   & A & 1.2230    & 2  & 81&   0.015 & 0.023 & 0.006 & 0.09& 0.001& 41.9& 66.9 & 16.8 & A17 \\
J2219+0229 & I & 0.98075 & 2 & 43 & 0.139 & 0.188 & 0.090 & -5.31 & -0.12 & 9.85 & 10.32 & 9.85 & D17b\\
J2245--3430 & A & 0.3562 & 1 & 39 & 0.128 & 0.128 & 0.128 & 1.18 & 0.031& 13.7 & 13.7 & 13.7 & A20\\
B2351+456 & I & 0.77945 & 1 & 118 & 0.228 & 0.228 & 0.228 & -5.64 & -0.048 & 60.3 & 60.3 & 60.3 & D04\\
\hline
\label{t1}
\end{tabular}
{References: L01 -- \citet{lb01}, M01 -- \citet{mot+01}, V03 -- \citet{vpt+03}, D04 -- \citet{dgh+04}, C06 -- \citet{cwm+06}, 
  G06 -- \citet{gss+06}, 
  J10 -- \citet{jbc+10}, C11 -- \citet{css11}, A15 -- \citet{asm+15}, Y16 -- \citet{ysd+16}, A17 -- \citet{akp+17}, D17a --  \citet{dsgj17}, 
  D17b --  \citet{dsg+17},  M17a -- \citet{mmo+17}, M17b -- \citet{mas+17}, O17 -- \citet{omd+17}, 
  A18a-- \citet{ak17}, A18b -- \citet{ak18},  J18 -- \citet{jgs18}, A19 -- \citet{adi19}, 
  D19 -- \citet{dsgj19}, G19 -- \citet{gam+19}, A20 -- \citet{asd+19}, C20 -- \citet{ckc20}, M20 -- \citet{mpg+20}, S20 -- \citet{sma+20}, G21 -- \citet{gss+21}.\\
  Note: 
  \citet{mas+17} report a detection at $z_{\text{abs}} = 0.42016$, although, due to the lack of optical data, it is unknown whether this is associated or intervening.}
\end{minipage}
\end{table*} 

\subsubsection{Extraction and processing of spectra}

\begin{figure}
\centering \includegraphics[angle=0,scale=0.65]{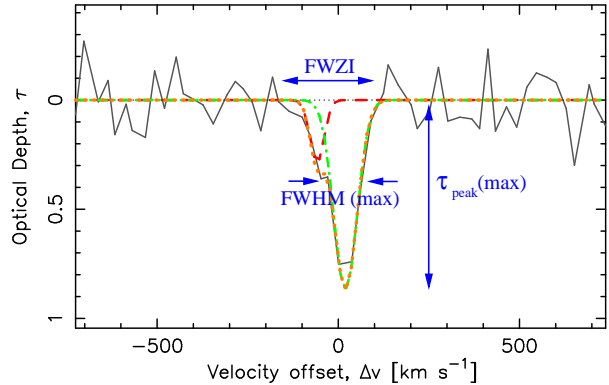}
\caption{Example of Gaussian fits to a relatively noisy, but optically thick, absorption spectrum \citep{ckc20}. The ordinate has been
normalised to optical depth and the abscissa to the velocity offset from the 
weighted mean value (see \citealt{cdda16}). The red and green fits show the individual Gaussian profiles and the orange fit the sum.}
\label{spec}
\end{figure}

Since the spectra were generally unavailable, as before, we obtained these from the literature by digitising the
appropriate figures.  This was done using ADS's {\em Dexter Data Extraction Applet} \citep{dae+01}. Upon extraction,
each spectrum was normalised by converting the ordinate to optical depth and the abscissa to velocity.  In order to
obtain the features required for the machine learning, Gaussians were fit to each spectrum using a custom written {\sf
  pgplot} wrapper for the {\sf python} {\tt curve\_fit} function, an example of which is shown in Fig.~\ref{spec}.
Since all of the spectra had already had the continuum fitted or removed, we left this as is.  The number of Gaussian
profiles was obtained by minimising the $\chi$-squared residual between the fit and the data where this led to a low
number of profiles, otherwise the minimum number which gave a residual consistent with the noise from the line free part
of the spectrum was used. In most cases the number of profiles required was in agreement with the published number.  As
per \citet{cdda16}, for the purpose of the FWZI, the end points of the profile where defined by where the compound fit
reached an optical depth of $\tau < 10^{-4}$.

\subsubsection{Features}

We use the same features as previously, listed in Table~\ref{t1} and shown in Fig.~\ref{spec}.  These are the number of
Gaussian fits to the profile, the span of the absorption and the width and depth information for the Gaussian
components.  As before, we also include the offset of the Gaussians from the weighted mean velocity over the whole
spectrum, which is defined by the velocity which contains half of the velocity integrated optical depth.

\begin{figure*}
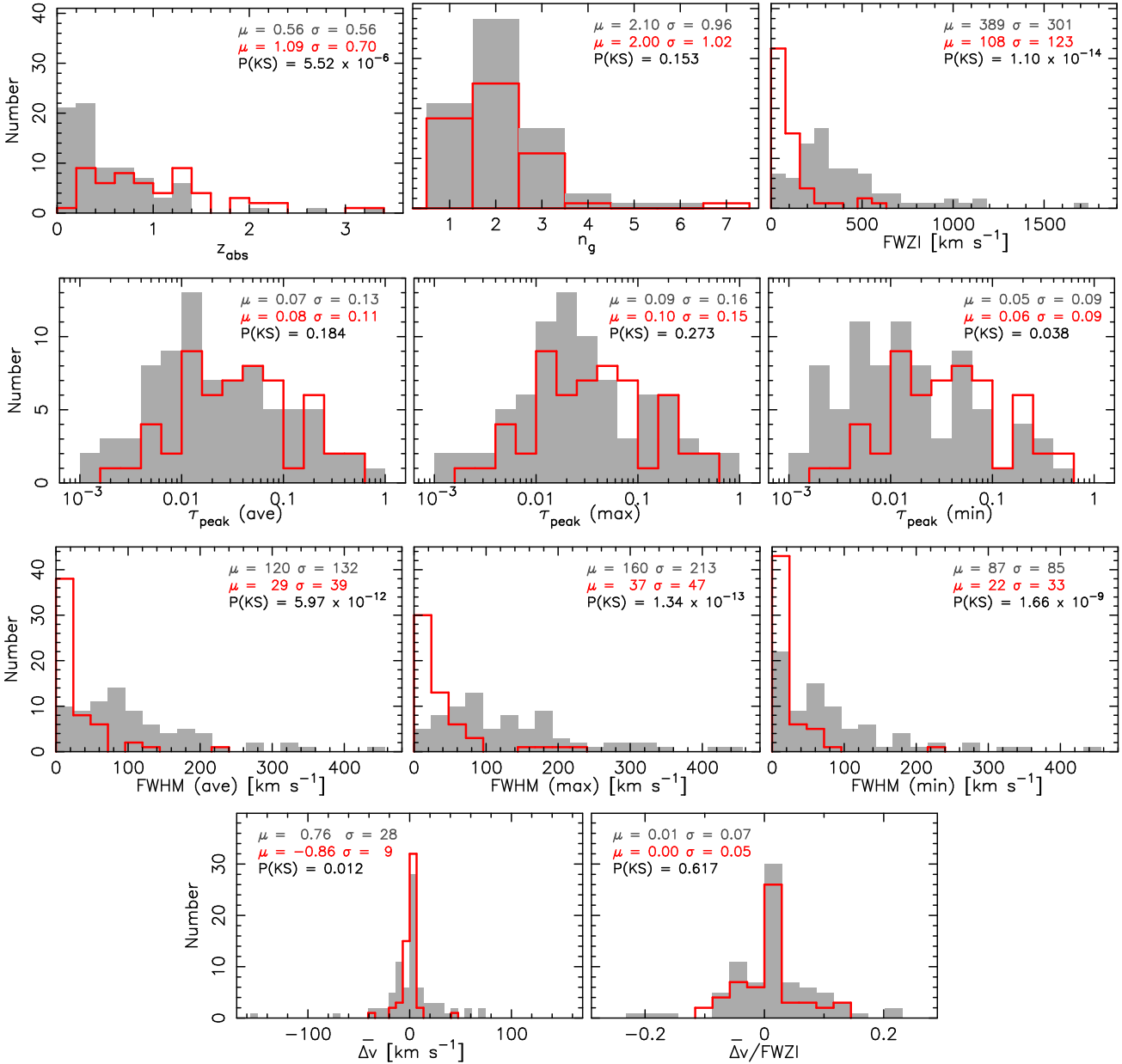

\centering \includegraphics[angle=-90,scale=0.52]{histo_z-bin=18.eps}
\vspace*{2mm}
\centering \includegraphics[angle=-90,scale=0.52]{histo_n_g-rect-bin=37.eps}
\centering \includegraphics[angle=-90,scale=0.52]{histo_FWZI-bin=24.eps}
\vspace*{2mm}
\centering \includegraphics[angle=-90,scale=0.52]{histo-tau_ave-bin=17.eps}
\centering \includegraphics[angle=-90,scale=0.52]{histo-tau_max-bin=17.eps}
\centering \includegraphics[angle=-90,scale=0.52]{histo-tau_min-bin=17.eps}
\vspace*{2mm}
\centering \includegraphics[angle=-90,scale=0.52]{histo-FWHM_ave-bin=20.eps}
\centering \includegraphics[angle=-90,scale=0.52]{histo-FWHM_max-bin=20.eps}
\centering \includegraphics[angle=-90,scale=0.52]{histo-FWHM_min-bin=20.eps}
\centering \includegraphics[angle=-90,scale=0.52]{histo-ave_v-bin=50.eps}
\centering \includegraphics[angle=-90,scale=0.52]{hist-ave_v_over_FWHHZ-bin=5.eps}
\caption{The distribution of features for the sample. The filled histogram represents the associated
  absorbers and the unfilled the intervening absorbers. In addition to the mean and standard
  deviation, we show the Kolmogorov-Smirnov probability that the associated and intervening
  absorbers are drawn from the same sample.}
\label{features}
\end{figure*}

The feature distributions are shown in Fig.~\ref{features}.
Apart from the redshift distributions, we see that the only real distinction is the velocity widths,
with clear differences in the FWZIs and FWHMs.  From the
observed optical depths, \citet{cdda16} suggest that  the additional nuclear component, hypothesised by
unified schemes of active galactic nuclei (AGN, e.g. \citealt{ant93,up95}), could contribute to the 
broader associated profiles. The other/additional possibility is that associated absorption is
preferably detected through the disk of the host (cf. \citealt{cw10}), whereas intervening absorption may be
preferably detected in face-on systems, where the coverage of the background continuum is
maximised \citep{cras16}.

Interestingly, there is no difference in the number of Gaussian fits between the two classes. This
was previously noted by \citeauthor{cdda16} and interpreted as being a consequence of the different
noise levels in the spectra diluting intrinsic differences in the profiles. The associated absorbers
also tend to be at lower redshift, due to the ionisation of the neutral gas by the active galactic
nucleus in $z\gapp1$ radio sources \citep{chj+19}, although both classes are subject to an
increasingly poor radio frequency environment as the observed frequency decreases from 1.4~GHz,
making high redshift detection difficult.\footnote{Note also that the detection rate of intervening
absorbers is also lower at $z_{\text{abs}}\gapp1$ due to the geometry effects of an
expanding Universe \citep{cur12}.}

\subsection{Machine Learning}

In our previous study we used {\sc weka} \citep{hfh+09}, a suite of machine learning algorithms, for
the classification of the spectra. However, the {\sf
  sklearn}\footnote{https://scikit-learn.org/stable/} module of python is more widely used by the
astronomy community (e.g. \citealt{vcig12}) and can be run from the command line.  This allows us to
run a number of trials for each algorithm, shuffling the training and validation data each time in
order to obtain a representative cross-validation score.

We test four types of  common classifiers -- {\em  logistic regression}, {\em k-nearest neighbour}, {\em support vector classifier}
and the {\em decision tree classifier}. Since there is a mismatch in the two binary classes (associated \& intervening),
we perform random under-sampling, where 56 associated absorbers are randomly selected for training and validation.
The data are then shuffled and the features normalised.
Given the small size of the data-set, we use 10-fold cross-validation to split the data into ten sets,
each of size $(56+56)/10 \approx 11$, of which ten are used to train the model and one to validate.
This is repeated ten times, each time shuffling the data, with the mean accuracy being reported. 
This process is run 1000 times and the mean score of each algorithm recorded.

 \subsubsection{Logistic Regression}

\begin{figure}
\centering \includegraphics[angle=-90,scale=0.48]{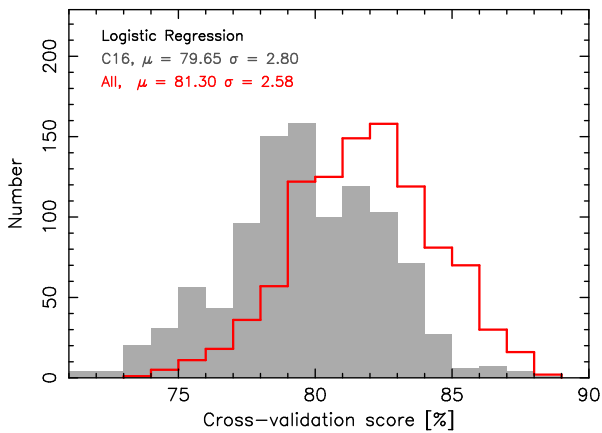}
\caption{The results of  1000 trials using logistic regression, showing the mean
score and standard deviation. The filled histogram represents the previous \citep{cdda16}
data and the unfilled histogram all of the data.}
\label{LR}
\end{figure}

 Logistic regression is analogous to multi-variable linear regression, but instead of a fit yields a binary
 result. Thus, it is particularly suited to this problem, where we are classifying an absorber as either associated or
 intervening.  The algorithm compresses a linear combination of several variables (features) with a logistics sigmoid to
 yield a value of between 0 and 1. For the binary model, the prediction is labelled with one of these two end values,
 depending upon its probability ($0\rightarrow1$), or odds ($0\rightarrow\infty$).\footnote{We chose 0 for intervening
   and 1 for associated absorption.}  For {\em even odds}, the values are 0.5 and 1, respectively. The algorithm has the
 advantage that is quicker than other machine learning techniques, although it is subject to the same errors as linear
 regression, such as skewing by outliers and the requirement of a sufficiently large sample.  Also, the presence of a
 feature which perfectly separates the two binary classes will prevent the feature weight from converging, with the
 infinite optimal weight halting the training. 
Using the default {\sf sklearn} values\footnote{https://scikit-learn.org/stable/modules/generated/\\~~~sklearn.linear\_model.LogisticRegression.html}, the logistic regression model  of all of the data we get a mean
 cross-validation score of 81.3\% (Fig.~\ref{LR}), which is
a $1.7$\% ($\approx0.6\sigma$) improvement over the previous data.

\subsubsection{k-Nearest Neighbour}
\label{sec:knn}

\begin{figure}
\centering \includegraphics[angle=-90,scale=0.48]{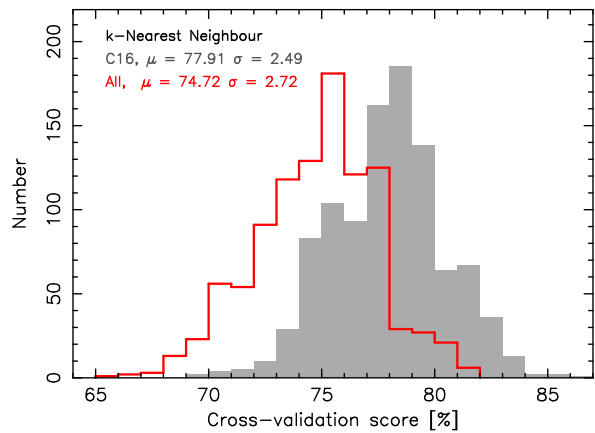}
\caption{As Fig.~\ref{LR}, but for the $k$-nearest neighbour algorithm.}
\label{kNN}
\end{figure}

The $k$-nearest neighbour (kNN) algorithm maps the variables to a feature space and then compares
the Euclidean distance between a test point and its $k$ nearest neighbours. It then assigns a
weighted combination of the target values with the nearest neighbours in order to place the test
object in a group. The kNN algorithm is relatively computationally expensive and, like logistic
regression, is sensitive to outliers. A further disadvantage is that irrelevant features can lead the
learning astray. For the kNN method  we are required to supply the number of neighbours to use and, from several
mini-trials (of 10), we found that $k\approx15$ was optimal, although anything in the range $k = 5
-20$ gave similar results.  The kNN algorithm has been used extensively in predictive modelling to
obtain photometric redshifts from source magnitudes
(e.g. \citealt{rws+01,wrs+04,mhp+12,hdzz16,cur20}).  In this case we use it for binary
classification, where it does not perform as well as the logistic regression.  Furthermore, the
additional data decreases the cross-validation score, resulting in a 
$\approx1.2\sigma$ worse performance (Fig.~\ref{kNN}).

\subsubsection{Support Vector Classifier}

\begin{figure}
\centering \includegraphics[angle=-90,scale=0.48]{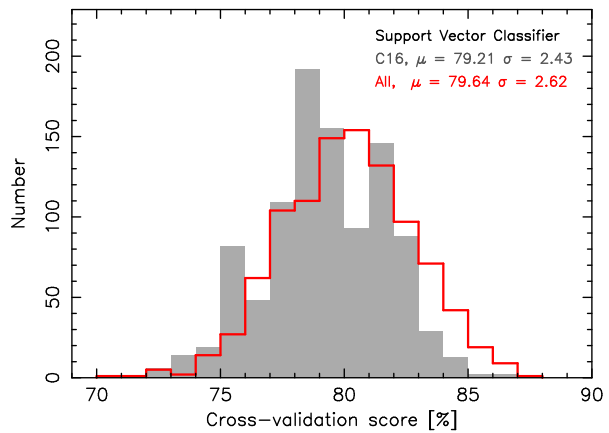}
\caption{As Fig.~\ref{LR}, but for the support vector classifier.}
\label{SVC}
\end{figure}

A support vector machine  constructs a hyperplane in a high dimensional space in order to perform classification,
regression and outlier detection. 
Support vectors are points that reside closest to the hyperplane and 
in binary classification the training maximises the distance between the two categories. Further data are
transformed into the same space and assigned a category based upon where in the space the are located.
Although the support vector machine classifier (SVC) is computationally fast, it is not suitable 
for noisy data (overlapping features) nor large data sets. This could limit the applicability of this
algorithm in classifying 21-cm absorption from large surveys (Sect.~\ref{intro}), although in this case the SVC 
is a reasonable performer. Using the  default {\sf sklearn} values\footnote{https://scikit-learn.org/stable/modules/generated/sklearn.svm.SVC.html}, we obtain 
a cross-validation score of $\approx79$\%  for both the previous and
the previous plus the additional data  (Fig.~\ref{SVC}).

\subsubsection{Decision Tree Classifier}

\begin{figure}
\centering \includegraphics[angle=-90,scale=0.48]{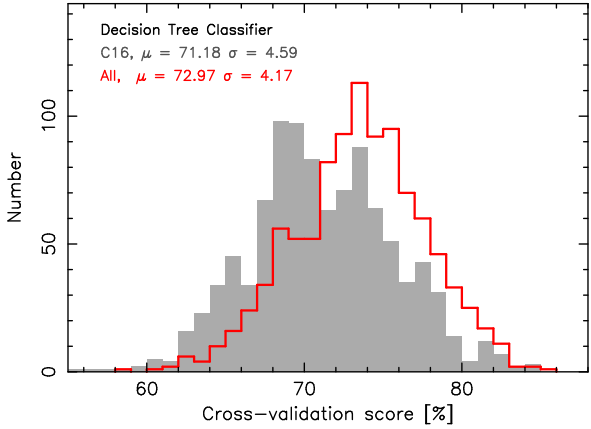}
\caption{As Fig.~\ref{LR}, but for the decision tree classifier.}
\label{DTC}
\end{figure}

Like the other algorithms, decision trees can be used for both classification and regression.  The algorithm builds a
classification model based upon a a tree structure, which branches the data-set (top node) into smaller subsets (child
nodes), according to a predefined decision boundary.  With one node on either side of the boundary, the process is
iterated through further branching until a predefined stopping criterion in reached. DTC has the advantage that it is
not as sensitive to outliers as some of the other algorithms, although the tree choices can be biased due to the
sequential nature of the algorithm.  Over-fitting can also be a problem, which can be mitigated by limiting the maximum
tree depth.  From several mini-trials, we found the maximum depth to make little difference to the score and so used the
default of no maximum depth, in which branches split until each ``leaf'' only contained a sample of two.  From the 1000
trials, we see that the decision tree classifier (DTC) is the poorest performing algorithm (Fig.~\ref{DTC}).

\section{Discussion}

\subsection{Results}

We see that the logistic regression is best performer, giving a cross-validation score of $\gapp80$\%, which
improves with the addition of the new data.  In Table~\ref{conf} we show the mean confusion matrix of the test results
\begin{table}
\centering
\caption{The mean confusion matrix from 1000 runs of the logistic regression.}
\begin{tabular}{@{}l  cccc  @{}}
                  & \multicolumn{2}{c}{Predicted} & \multicolumn{2}{c}{Predicted} \\
                        & \multicolumn{2}{c}{associated} & \multicolumn{2}{c}{intervening} \\
                        & $\mu$& $\sigma$  &   $\mu$& $\sigma$  \\
Actual associated & 4.67 &   1.65& 1.55 & 1.19 \\
Actual intervening & 1.27&  1.12  & 6.52 & 1.78\\
\end{tabular}
\label{conf}  
\end{table} 
obtained from 1000 runs of the logistic regression. That is, testing on data unseen by the algorithm gives a mean of 
of 4.67 true positives (associated) and 6.52 true negatives (intervening), compared to 1.27 false positives 
and 1.55 false negatives. This gives a test score of  (4.67 + 6.52)/(4.67 + 6.52  + 1.27 + 1.55) = 0.80, which is close
to the cross-validation score.

The support vector classifier is the next best performer, with a mean score of $\approx79$\%, which also improves with
the addition of the new data. Following this  is the $k$-nearest neighbour algorithm, but in this case the additional data
significantly decreases the cross-validation score making  this arguably the worst performer. Last,
the decision tree classifier gives a mean score of  $\approx73$\%. Unlike the kNN, the DTC score increases
with the new data, although the spread in values is wide ($\sigma\gapp4$\%).

\subsection{Feature importance}

Previously, we found the FWZI and FWHM to be the most important features and, while we can 
extract the coefficients of the logistic regression, these are given in terms of the odds ratio and are therefore not
straightforward to interpret.  We therefore show the feature importance for the other algorithms (Table \ref{t2}), 
\begin{table}  
\centering
 \caption{The normalised importance of each feature for the $k$-nearest neighbour, support vector classifier and decision tree classifier. The means and standard deviations are obtained from 1000 trials of each.}
\begin{tabular}{@{}l  ccc  ccc ccc  @{}}
\hline
\smallskip         %    ./run_others.csh
               & \multicolumn{2}{c}{kNN}  & \multicolumn{2}{c}{SVC} & \multicolumn{2}{c}{DTC} \\
Feature & $\mu$& $\sigma$  &   $\mu$& $\sigma$  & $\mu$& $\sigma$  \\
\hline
$z_{\text{abs}}$ &  0 &  0.0005 &   0& 0&   0.198 & 0.045\\ %1
$n_{\rm g}$    & 0 &  0.001 &   0 &  0&   0.009 & 0.013\\ %2
FWZI              & 0. 254 & 0.024 & 0.260 &   0.030& 0.281 &   0.067\\ %3
$\tau_{\rm peak}$ (ave)& 0 &  0 &   0 &  0 &   0.056 & 0.045 \\ %4
$\tau_{\rm peak}$ (max)& 0 &  0 &  0 &  0 &   0.057 & 0.051 \\ %5
$\tau_{\rm peak}$ (min)&0 &  0 & 0 &  0 & 0.039 & 0.043\\  %6
$\overline{\Delta v}$  & 0 & 0.004 & 0 & 0.001 & 0.025 & 0.016\\ %7
$\overline{\Delta v}$/FWZI & 0&  0 & 0 &  0 &0.046 & 0.043\\  %8
FWHM (ave) &0.005 & 0.009 & 0.013 & 0.007 & 0.028 & 0.012\\ %9 
FWHM (max)  & 0.010 & 0.011 & 0.021 & 0.010 & 0.073 & 0.079\\ %10
FWHM (min) & 0.007 & 0.007 & 0.009 & 0.006 & 0.052 & 0.037\\  %11
\hline
\end{tabular}
\label{t2}  
\end{table} 
from which we see that the absorption redshift, number of Gaussian fits and the peak optical depth make zero
contribution to the kNN and SVC algorithms and where these are non-zero, for the DTC algorithm, 
the values are often dominated by the scatter. While the FWHM makes a minor contribution, as noted by \citet{cdda16},
the most important feature is the full line width. 

\begin{figure}
\centering \includegraphics[angle=-90,scale=0.48]{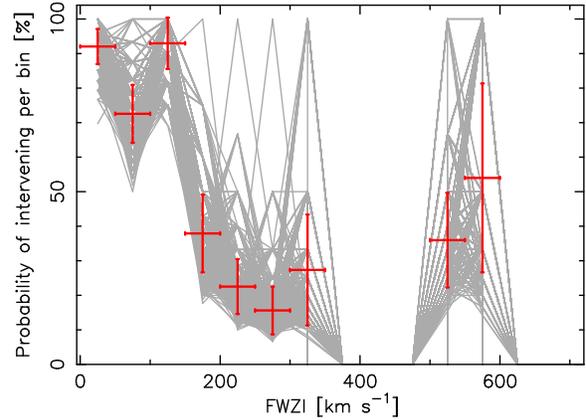}
\caption{The probability of the absorber being intervening based upon the FWZI of the profile (cf. Fig.~\ref{features}) in 50~\kms\ bins.  The traces show each of the 1000 trails, where for each we randomly select 56 of the associated systems.
The error bars show $\pm1\sigma$ about the mean binned value.}
\label{prob}
\end{figure}

This suggests the possibility of obtaining a probable absorber class
from the FWZI alone. Fig.~\ref{prob} shows the probability of the
absorber being intervening for a given FWZI.
From this, we see that if FWZI~$\leq50$~\kms\ there is a $\approx90$\% probability of the absorber being
intervening, with even odds, on average,  being reached between the 100 -- 150 ~\kms\ bins. 
Although subject to small numbers, this demonstrates that the type of absorption can only
be predicted with $\gapp70$\% accuracy from the FWZI where this is $\lapp200$~\kms.

\begin{figure*}
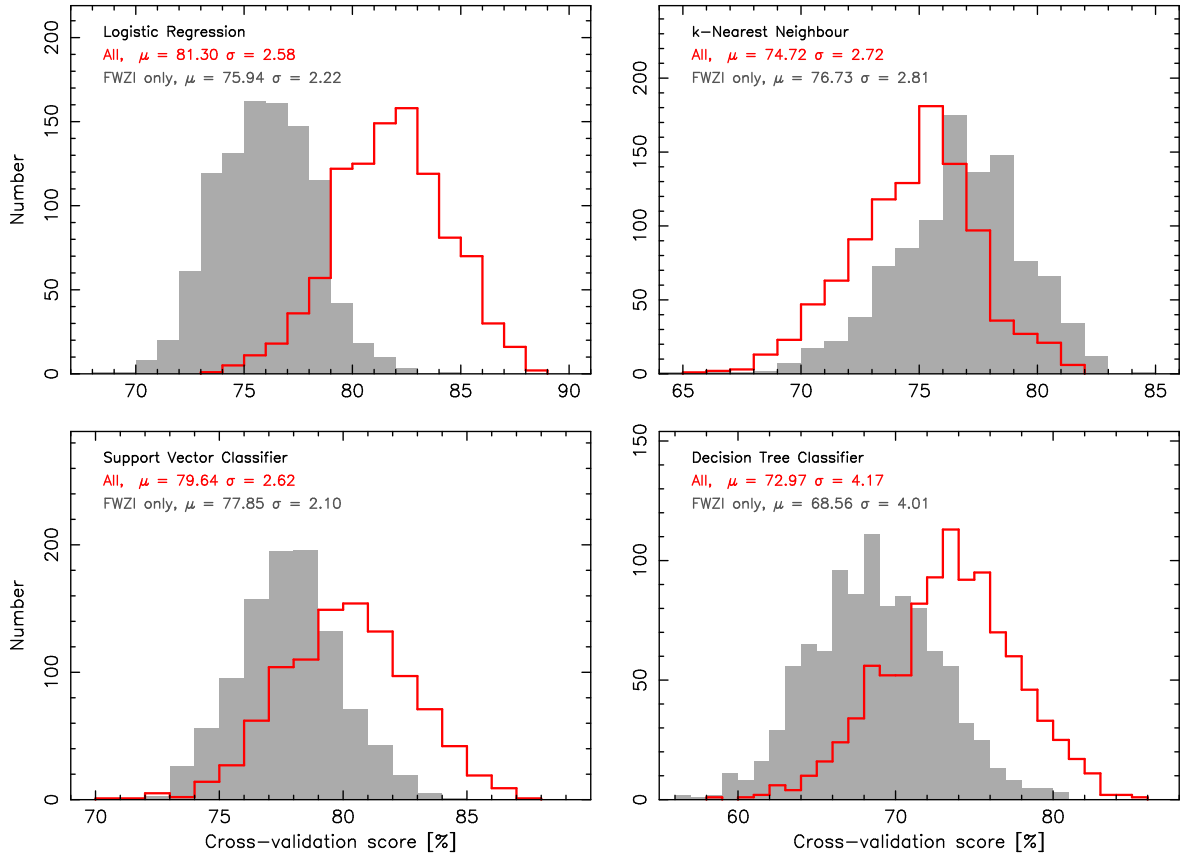

\centering \includegraphics[angle=-90,scale=0.48]{LR_feat_histo-bin=24.eps}
\centering \includegraphics[angle=-90,scale=0.48]{kNN_feat_histo-bin=22.eps}
\centering \includegraphics[angle=-90,scale=0.48]{SVC_feat_histo-bin=21.eps}
\centering \includegraphics[angle=-90,scale=0.48]{DTC_feat_histo-bin=33.eps}
\caption{Comparison between machine learning with all features (unfilled histogram)
and FWZI only (filled histogram) over 1000 trials.}
\label{feat_hist}
\end{figure*}

In order to further quantify the contribution of the other features, in Fig.~\ref{feat_hist} we show the algorithm performances
using only the FWZI as a feature.
From this, we see that the logistic regression exhibits the greatest improvement ($\approx2\sigma$), with support vector
classifier and decision tree classifier showing some to a  lesser degree. Again, for the $k$-nearest neighbour
algorithm the score decreases, although by a small amount with the binned modes being very close.  This suggest that the
LR algorithm makes most use of all the features and the kNN the least.  The degradation in kNN score with the addition of
the new data is therefore probably due to the introduction of noise from  the other features as well as the new data
not necessarily following a straightforward absorber type--FWZI relation: For the associated absorbers in the previous
data $\overline{\text{FWZI}} = 416$~\kms, cf.  389~\kms\ for all of the data.  Since the new data moves the mean of the FWZI 
of the associated absorbers in the opposite sense to that expected, if basing the classification on the  FWZI alone,
we expect  the algorithm to perform more poorly.\footnote{For the intervening absorbers, 
$\overline{\text{FWZI}} = 118$~\kms\ for the previous data, cf. 108~\kms\ for all of the data.}

\subsection{The absorption towards PKS\,1657--298}

As stated in Table~\ref{t1}, because of the lack of an optical spectrum the nature of the $z_{\text{abs}} = 0.42016$
absorption towards PKS\,1657--298 is unknown \citep{mas+17}. We can, however, find the likely absorber type through a
two-pronged approach: 1) prediction of the type from our logistic regression model and, 2) using machine learning to
obtain the redshift of the background source. As described in Sect.~\ref{intro}, the latter can be obtained from a
neural network trained on SDSS QSOs and then confidently applied to a radio selected source \citep{cmp21}.  The best
results are obtained when the photometry from nine bands, spanning the far-ultraviolet to near-infrared
($FUV,NUV,u,g,r,i,z,W1,W2$), are used. Scraping the photometry from the {\em NASA/IPAC Extragalactic Database}, the {\em
  Wide-Field Infrared Survey Explorer}, the {\em Two Micron All Sky Survey} (\citealt{scs+06}) and {\em Galaxy Evolution
  Explorer} (\citealt{bst17}) databases, as described in \citet{cmp21}, we find the photometry to be limited to
$\nu\leq20$~GHz apart from one X-ray measurement. The lack of photometry, therefore, does not permit a photometric
redshift determination.

Running 1000 trials to predict the absorber type from our logistic regression model, in 970 cases the absorption is
classed as associated.  Although, as mentioned above, there is no infrared photometry for PKS\,1657--298, \citet{mas+17}
suggest that the measured $K=14.9\pm0.1$ from 2MASS\,J17011004--2954423 may arise from PKS\,1657--298, from which it is offset by 4.3''. They then use the $K$-band magnitude--redshift relation (\citealt{dvs+02,wrjb03}) to obtain an emission
redshift similar to that of the 21-cm absorption ($z_{\text{em}} \approx0.42$), thus classifying the absorption as associated.
However, this method of estimating the redshift is only effective for (radio) galaxies, providing, at best, a lower
limit for quasar redshifts \citep{cm19}. No AGN classification could be found for this object, although the observed
X-ray flux of $4.9\times10^{-13}$~ergs~cm$^{-2}$~s$^{-1}$, obtained from archival XMM-Newton data \citep{mas+17},
\begin{figure}
\centering \includegraphics[angle=-90,scale=0.48]{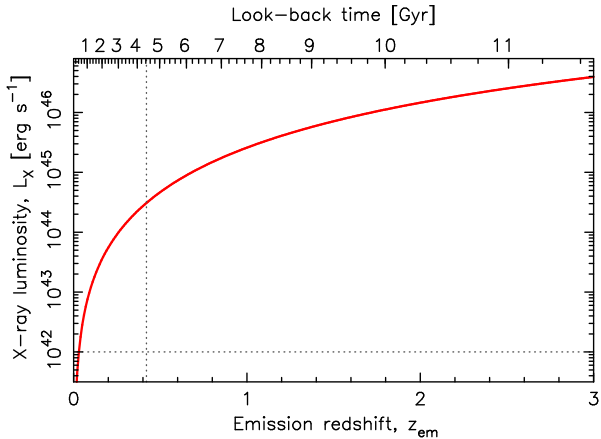}
\caption{The X-ray luminosity of PKS\,1657--298  as a function of redshift.  The dotted vertical line designates the absorption redshift and the horizontal line designates the lower limit for an AGN \citep{psv+12}.}
\label{X-lum}
\end{figure}
gives a luminosity is $L_{\text{X}} \gapp 5\times10^{44}$ ~ergs~s$^{-1}$ at $z\geq0.42$ (Fig.~\ref{X-lum}).\footnote{NED lists a 2--10~keV flux of $7.68\times10^{-13}$~ergs~cm$^{-2}$~s$^{-1}$, which is due to a Galactic low-mass X-ray
binary \citep{ver01,vig+07}.}  This is
relatively high \citep{gab+15}, in the range where the source is likely to be a type-1 AGN
\citep{hphk05,jdas11,ssr11}.  
This  is also in the range where star formation is believed to be suppressed by the AGN activity (\citealt{psv+12}, but see
\citealt{cd19}), thus suggesting that the source a quasar rather than a radio galaxy.

\section{Conclusions}

Being able to classify a redshifted \HI\ 21-cm spectrum as intervening or associated 
without the need for an optical spectrum would provide a powerful diagnostic for
detections  with the next generation of large radio telescopes. Previously \citep{cdda16}, from a sample of 
92 $z_{\text{abs}}>0.1$ systems, of which 48 were associated and 44 were intervening, we found that machine
learning algorithms could predict the absorber type with an $\approx80$\% accuracy. 
Updating the sample with the addition of 44 absorption systems not previously included, we run four
separate classifiers 1000 times in order to obtain a typical cross-validation score. We find that three of these
give a slight improvement in the score over the previous sample, with the fourth worsening. In decreasing order of score:
\begin{itemize}
\item[--]  Logistic regression gives a mean score of 81.3\%, compared to 79.7\% for the previous
sample.
\item[--]  Support vector classification gives a mean score of 79.6\%, compared to 79.2\%.
\item[--] $k$-Nearest neighbour gives a mean score of 74.2\%, compared to 79.9\%.
\item[--] Decision tree classification gives a mean score of 73.0\%, compared to 71.2\%.
\end{itemize}

For all algorithms we find the line-widths to be the most important features, particularly the full-width at zero
intensity, although, except for the kNN, this on its own does not perform as well as the full feature complement.  This
is most pronounced for the highest scoring algorithm (LR), while for the  kNN
algorithm, which exhibits a decrease in accuracy upon the addition of the new data, the inclusion of the other features
makes little difference to using the FWZI alone. The presence of ``irrelevant features'' leading the learning astray
is a known issue with $k$-nearest neighbour.  We therefore believe that the kNN is assigning most of the weighting to the
FWZI and, given that the mean FWZI of the associated absorbers in the whole sample is lower than the previous sample,
this leads to a poorer performance.

We use the logistic regression model to classify the $z_{\text{abs}}=0.42$ absorption towards PKS\,1657--298 which, in
the absence of an optical spectrum, is unknown \citep{mas+17}. Out of 1000 trials, the absorption is flagged as
associated in 970 cases, which is consistent with \citeauthor{mas+17}, who, from the  $K$-band magnitude--redshift
correlation for galaxies, estimate a source redshift of $z_{\text{em}} \approx z_{\text{abs}}$. However, this assumes
the $K$-band magnitude of a near-by source and the X-ray luminosity of $L_{\text{X}} \gapp 5\times10^{44}$ ~ergs~s$^{-1}$
suggests that the continuum source is a type-1 object (quasar), for which the $K$-band magnitude--redshift
relation does not hold \citep{cm19}. Thus, although, we also find the absorption to be associated, there is 
some uncertainty in the \citeauthor{mas+17} classification.
%Furthermore,  it is not clear whether this X-ray 

With this increase in sample size, giving what are still very limited numbers, logistic regression appears
to be the best performer and also the algorithm which makes most use of all of the spectral features.
This is also the case, to a lesser extent, for the support vector classification. The decision tree
classifier also improves with the addition of the new data, but with a mean score of $73$\% this
falls short of the LR and SVC performance ($\approx80$\%). Although more accurate at $\approx75$\%,
the kNN algorithm appears to be basing its predictions on the FWZI alone, which results
in a degradation of score with the addition of the new data. Based upon this small sample,
it appears higher accuracies maybe be achievable with 
the logistic regression algorithm trained upon  appropriately sized samples.
With the addition of more training data over the next few years, this could provide an important diagnostic for forthcoming
surveys with the Square Kilometre Array.

\section*{Acknowledgements}

I wish to thank the referee, Jeremy Darling, for his very helpful comments. 
This research has made use of the NASA/IPAC Extragalactic Database (NED) which is operated by the Jet Propulsion Laboratory, California
Institute of Technology, under contract with the National Aeronautics and Space Administration and NASA's Astrophysics
Data System Bibliographic Service. This research has also made use of NASA's Astrophysics Data System Bibliographic Service.

\section*{Data availability} 

The data underlying this article are available in its online supplementary material.

\label{lastpage}

\end{document}